\documentclass[final,5p,times,fixfloat]{elsarticle}

\biboptions{numbers,sort&compress} %

\usepackage{amsmath,graphicx,verbatim,epsfig,amssymb,dsfont}
\usepackage{epstopdf}
\usepackage[latin1]{inputenc}
\usepackage{times}
\usepackage{tabularx}

\usepackage{txfonts}
\usepackage{lineno}

\journal{Physics Letters B}

 \usepackage{ifpdf}
 \ifpdf
 \usepackage[pdftex]{hyperref}
 \else
 \usepackage[hypertex]{hyperref}
 \fi

 \hypersetup{
   pdftitle={},%
   pdfauthor={},%
   pdfsubject={},%
   pdfkeywords={},%
   pdfstartview={},%
   bookmarksopen=true, breaklinks=true, debug=true, %
   colorlinks=true, linkcolor=red, citecolor=blue, urlcolor=blue
 }

\usepackage{float} 
\usepackage[caption=false]{subfig} 

\newcount\savefnused
\newcount\savefndone

\newcommand{\savefootnote}[2][\empty]
{\ifx\empty#1\footnotemark\else\footnotemark[#1]\fi
 \global\advance\savefnused by 1
 \expandafter\xdef\csname savefnmark\the\savefnused\endcsname{\thefootnote}%
 \expandafter\xdef\csname savefntext\the\savefnused\endcsname{#2}%
}
\newcommand{\flushfootnote}{\loop\ifnum\savefndone<\savefnused
  \global\advance\savefndone by 1
  \footnotetext[\csname savefnmark\the\savefndone\endcsname]%
    {\csname savefntext\the\savefndone\endcsname}%
  \global\expandafter\let\csname savefnmark\the\savefndone\endcsname\relax
  \global\expandafter\let\csname savefntext\the\savefndone\endcsname\relax
\repeat}

\usepackage{multirow,bigdelim}

\usepackage{textcomp}
\usepackage{comment}

\usepackage{lineno}

\usepackage{float} 
\usepackage{subfig} 
\usepackage{amsmath} %
\usepackage{amstext}

\usepackage{graphicx}%
\usepackage{tabularx,ragged2e}%

\newcolumntype{Y}{>{\centering\arraybackslash}X}

\newcommand{\Ga}{\Gamma}

\newcommand{\nn}{\nonumber}

\newcommand{\bn}{{\bar n}}

\newcommand{\be}{\begin{equation}}
\newcommand{\ee}{\end{equation}}
\newcommand{\bea}{\begin{eqnarray}}
\newcommand{\eea}{\end{eqnarray}}
\newcommand{\balign}{\begin{align}}
\newcommand{\ealign}{\end{align}}
\newcommand{\as}{\alpha_s}

\newcommand{\cd}{\!\cdot\!}

\renewcommand{\d}{\mathrm{d}}

\newcommand{\sandwich}[3]{\left< #1 \right | #2 \left | #3 \right >}

\newcommand{\bg}{\begin{gather}}
\newcommand{\foma}{\end{gather}}

\newcommand{\noopsort}[1]{}

\newcommand{\vecb}[1]{\mbox{\boldmath $#1$}}
\newcommand{\vecbe}[1]{\mbox{\boldmath ${\scriptstyle #1}$}}
\newcommand{\vecbp}[1]{\mbox{\boldmath $#1_\perp$}}

\def\L{\Lambda}

\def\z{\zeta}

\def\<{\langle}
\def\>{\rangle}

\def\a{\alpha}
\def\b{\beta}
\def\g{\gamma}  \def\G{\Ga}
\def\d{\delta}

\def\x{\xi}

\def\m{\mu}

\def\z{\zeta}

\def\({\left(}
\def\[{\left[}
\def\){\right)}
\def\]{\right]}

\def\ln{\hbox{ln}}

\def\bnslash{\bar n\!\!\!\slash}

\def\le{\left }
\def\ri{\right}

\def\lqcd{\L_{\rm QCD}}

\newcommand{\ben}{\begin{eqnarray}}
\newcommand{\een}{\end{eqnarray}}

\newcommand{\bef}{\begin{figure}[htb]\centering}
\newcommand{\eef}{\end{figure}}

\usepackage[normalem]{ulem} 

\begin{document} 
\begin{frontmatter}

\title{QCD$\times$QED evolution of TMDs}
\author[a,b]{Alessandro Bacchetta}
\ead{alessandro.bacchetta@unipv.it}
\author[b]{Miguel G. Echevarria}
\ead{mgechevarria@pv.infn.it}

\address[a]{Dipartimento di Fisica, Università degli Studi di Pavia, via Bassi 6, 27100 Pavia, Italy}
\address[b]{Istituto Nazionale di Fisica Nucleare, Sezione di Pavia, via Bassi 6, 27100 Pavia, Italy}

\begin{abstract}

{\small
We consider for the first time the QED corrections to the evolution of (un)polarized quark and gluon transverse-momentum-dependent distribution and fragmentation functions (TMDs in general).
By extending their operator definition to QCD$\times$QED, we provide the mixed new anomalous dimensions up to ${\cal O}(\alpha_s\alpha)$ and the pure QED ones up to ${\cal O}(\alpha^2)$.
These new corrections are universal for all TMDs up to the flavor of the considered parton, i.e., the full flavor universality of TMD evolution found in pure QCD is broken in QCD$\times$QED by the presence of the electric charge.
In addition, we provide the leading-order QED corrections to the matching coefficients of the unpolarized quark TMD parton distribution function onto its integrated counterparts at ${\cal O}(\as^0\a)$.
}
\end{abstract}

%
\end{frontmatter}
%

\section{Motivation}
\label{sec:motivation}

Transverse-momentum-dependent parton distribution and fragmentation functions
(TMDs in general) encode the 3-dimensional dynamics of partons in momentum space (see,
e.g., Refs.~\cite{Angeles-Martinez:2015sea,Rogers:2015sqa,Diehl:2015uka} for recent reviews). 

In the last years, TMD factorization, universality, and evolution properties have been placed on a firm theoretical ground, thanks to important contributions from several groups (see, e.g.,
Refs.~\cite{Ji:2004wu,Collins:2007nk,Rogers:2010dm,Becher:2010tm,Chiu:2012ir,Mantry:2010bi,Collins:2011zzd,Aybat:2011zv,GarciaEchevarria:2011rb,Echevarria:2012js,Vladimirov:2017ksc}).
Considerable progress has been made in the perturbative calculation of the QCD evolution of TMDs~ \cite{Aybat:2011zv,Echevarria:2012pw,Echevarria:2014rua,Echevarria:2015byo,Li:2016ctv,Vladimirov:2016dll}, as well as their matching onto the corresponding integrated Parton Distribution Functions (PDFs) or Fragmentation Functions (see, e.g., Refs.~\cite{Catani:2013tia,Gehrmann:2014yya,Echevarria:2016scs,Bacchetta:2013pqa,Gutierrez-Reyes:2017glx,Buffing:2017mqm,Gutierrez-Reyes:2018iod}).   
In the present article, we take into consideration for the first time also QED corrections to TMDs.

From the phenomenological side, a good amount of unpolarized and polarized data from different hadronic processes is already available and has been used for the extraction of TMDs with QCD
evolution~\cite{Echevarria:2014xaa,DAlesio:2014mrz,Bacchetta:2017gcc,Scimemi:2017etj,Aybat:2011ta,Anselmino:2012aa,Kang:2015msa}. 
To describe the data, especially at colliders, it is essential to include evolution effects. 
In general terms, data at low transverse momentum ($q_T^2 \approx \lqcd^2$) are strongly affected by nonperturbative contributions, which are at the moment not precisely constrained. 
At intermediate transverse momentum ($\lqcd^2 \ll q_T^2 \ll Q^2$), the perturbative contributions entailed in TMD evolution play a dominant role: precision studies in this region require a detailed knowledge of the nonperturbative components of the TMDs, as well as the best possible knowledge of all contributions to the evolution of TMDs. 
Data from present hadron-hadron colliders (see, e.g., Refs.~\cite{Aad:2014xaa,Aad:2015auj,Aaij:2015gna,Aaij:2015zlq,Aaij:2016mgv,Khachatryan:2016nbe}) already call for the highest possible theoretical precision. 
This need will become even more pressing with future high-precision experiments~\cite{Accardi:2012qut,Dudek:2012vr,Brodsky:2012vg,Kikola:2017hnp,Hadjidakis:2018ifr}.

In this context, it is necessary and timely to consider the QCD$\times$QED contributions to TMD evolution. 
Similar corrections have been already taken into consideration for observables involving collinear PDFs~\cite{Roth:2004ti,Martin:2004dh,Ball:2013hta,deFlorian:2015ujt,deFlorian:2016gvk,Mottaghizadeh:2017vef,deFlorian:2018wcj}. 
Shortly before the completion of our work, similar improvements have been suggested also for the description of the transverse-momentum spectrum of $Z$ bosons produced at hadron colliders by Cieri-Ferrera-Sborlini~\cite{Cieri:2018sfk}:
their results are intimately connected to ours, but we adopt a complementary approach, based on the TMD framework, which makes it easy, among other things, to generalize our results also to polarized TMDs and fragmentation functions.

The paper is organized as follows. 
We will first extend the current definition of TMDs in QCD to QCD$\times$QED.
Then we will focus on the TMD evolution kernel, and provide the mixed corrections to the relevant anomalous dimensions up to ${\cal O}(\as\a)$ and the pure QED ones up to ${\cal O}(\a^2)$. 
Finally, we will consider the unpolarized quark TMDPDF and provide the leading-order QED corrections to the matching coefficients onto its integrated counterparts at ${\cal O}(\a)$.

\section{(Un)polarized quark/gluon TMDs in QCD$\times$QED}
\label{sec:definition}

In this section we extend the known definition of quark/gluon TMDs given in QCD \cite{GarciaEchevarria:2011rb,Echevarria:2012js,Collins:2011zzd} to QCD$\times$QED.
We first focus on quark TMDs, taking for simplicity the unpolarized quark TMDPDF, since all the considerations can be straightforwardly be extended to all the other polarized TMDPDFs and (un)polarized TMDFFs.
Then at the end of the section we will focus on gluon TMDs.

In simple terms, and leaving aside the details about the infrared/rapidity regulators which can be found in the mentioned references, the unpolarized $i$-flavor quark TMDPDF is defined as:
\begin{align}
\label{eq:TMDPDF}
f_{i/P}(x,k_{T}) &=
\int d^2\vecb b_\perp\, e^{i\vecbe b_\perp \cdot \vecbe k_{\perp}}\,
\tilde{J}_{i/P}(x,b_T)\,
\sqrt{\tilde{S}_i(b_T)}
\,,
\end{align}
where $\tilde J_{i/P}$ is the collinear matrix element and $\tilde S_i$ the soft function.
The twiddle refers to coordinate space.

Factorization in QCD$\times$QED follows the same logic as in pure QCD, so we can easily incorporate the QED contributions to the above matrix elements, which should now be given by:~\footnote{We use light-cone vectors $n$ and $\bn$ with $n^2=\bn^2=0,\; n\cdot\bn=2$.
A generic vector $v$ is decomposed as $v^\m=v^+n^\m/2+v^-\bn^\m/2+v_\perp^\m$ with $v^+=\bn\cd v$ and $v^-=n\cd v$. We denote $v_T=|v_\perp^\m|$.}
\begin{align}
\label{eq:JandS}
&J_{i/P}(x,k_{nT}) = 
\frac{1}{2} \int\frac{dy^-d^2\vecb y_\perp}{(2\pi)^3}
e^{-i(\frac{1}{2}y^-xP^+-\vecbe y _\perp \cdot \vecbe k _{n\perp})}
\nn\\
&\times
\frac{1}{2}\sum_{S}
\sandwich{PS}{\le[\bar\x_{n} W^T_n \widehat{W}^T_{i,n}\ri](0^+,y^-,\vecbp y)
\frac{\bnslash}{2}
\le[\widehat{W}^{T\dagger}_{i,n} W_n^{T\dagger} \x_{n}\ri](0)}{PS}
\,,
\nn\\
&S_i(k_{sT}) =
\int\frac{d^2\vecb y_\perp}{(2\pi)^2}
e^{i\vecbe y _\perp \cdot \vecbe k _{s\perp}}
\nn\\
&\times
\frac{{\rm Tr}_c}{N_c}
\sandwich{0}{\le[S_n^{T\dagger} S_\bn^T
\widehat{S}_{i,n}^{T\dagger}
\widehat{S}_{i,\bn}^T \ri]
(0^+,0^-,\vecb y_\perp)\le[S^{T\dagger}_\bn S_n^T
\widehat{S}^{T\dagger}_{i,\bn}
\widehat{S}_{i,n}^T \ri](0)}
{0}
\,.
\end{align}
The newly introduced QED Wilson lines, denoted by a hat on top, are the same as their QCD analogues, but with the gluon field replaced by a photon field.
Moreover, the path ordering does not apply in their case, since QED is an Abelian theory.
The exact definitions of the collinear and soft Wilson lines depend on the considered process, which will make them either past- or future-pointing.
We note that the newly introduced QED Wilson lines should follow, for a given process, the same direction as their corresponding QCD analogues.
Explicit expressions can be found e.g. in \cite{GarciaEchevarria:2011rb,Echevarria:2014rua}.
For instance, for Drell-Yan kinematics, we will have the following collinear gluon Wilson line and its corresponding collinear photon Wilson line (see e.g.~\cite{GarciaEchevarria:2011rb}):
\begin{align}
W_{n}(x) &=
\bar P\exp\Big[
ig_s\int_{-\infty}^{0}ds\, \bn\cdot A_{n}(x+s\bn)
\Big]
\,,
\nn\\
\widehat{W}_{i,n}(x) &=
\exp\Big[
ieQ_i\int_{-\infty}^{0}ds\, \bn\cdot B_{n}(x+s\bn)
\Big]
\,,
\end{align}
and similarly for the rest.
While $A_n$ stands for a $n$-collinear gluon, $B_n$ stands for a $n$-collinear photon.
Notice that the collinear photon Wilson line, as well as the soft photon ones, depends on the quark flavor through the charge $eQ_i$ (with $Q_i$ the fractional charge, i.e., $Q_u=2/3$, $Q_d=-1/3$ and so on).
The newly introduced QED Wilson lines guarantee that the collinear and soft matrix elements above are also QED gauge invariant, in addition to QCD gauge invariance.
The superscript $T$ denotes the necessity to include transverse gauge links to maintain the gauge invariance of the matrix elements for any gauge~\cite{GarciaEchevarria:2011md}.

The provided new definition of the unpolarized quark TMDPDF in QCD$\times$QED can be extended in a similar manner to all the other (un)polarized quark TMDs, both distribution and fragmentation functions.
To do so, one just needs to consider the proper polarizations and hadronic in/out states, on top of the addition of the soft/collinear photon Wilson lines introduced above.

In the case of (un)polarized gluon TMDs (see e.g. \cite{Echevarria:2015uaa}), the unsubtracted gluon correlator $J_{g/P}$ (analogous to $J_{i/P}$ in \eqref{eq:JandS}) does not acquire any photon Wilson lines, since gluons do not have electromagnetic charge and thus do not couple to photons.
In other words, one can say that the bi-local gluon-gluon correlator that appears in $J_{g/P}$ is already QED gauge invariant.
This fact has its impact as well on the extension of the soft function for gluon TMDs from QCD to QCD$\times$QED.
In this case, no soft photon Wilson lines are needed either, so the soft function for gluon TMDs in QCD$\times$QED remains the same as in pure QCD.
This can be understood as well since the absence of collinear photon Wilson lines in $J_{g/P}$ will not create any rapidity divergences which need to be cancelled by the corresponding soft photon Wilson lines in the soft factor.
In any case, gluon TMDs will still receive QED corrections from higher-order diagrams, in which fermion loops arise and thus allow for photons to appear as well.
Indeed, the leading-order QED corrections to the evolution kernel of gluon TMDs appear at ${\cal O}(\as\a)$, as we show below.

\section{QED corrections to the TMD evolution kernel}
\label{sec:evolution}

The TMDs in QCD depend on two scales: the factorization scale $\mu$ and the rapidity scale $\z$ (which is related to the arbitrary rapidity cutoff used to separate the soft function into two pieces).
Thus, their evolution kernel is such that it connects these two scales between their initial and final values.
In this section we provide the mixed QCD$\times$QED corrections to the anomalous dimensions that build the TMD evolution kernel at ${\cal O}(\as\a)$ and the pure QED ones up to ${\cal O}(\a^2)$.
We anticipate that these corrections will be universal for all (un)polarized quark and gluon TMD parton distribution and fragmentation functions, although they will depend on the flavor of the involved parton.

The renormalization group equations for the (un)polarized quark and gluon TMDs in QCD$\times$QED are (see e.g. \cite{Echevarria:2012pw,Echevarria:2014rua} for the pure QCD case):
\begin{align}
\label{eq:quarkTMDevo}
&\frac{d}{d\ln\m}\ln{\tilde F_i}(x,b_T;\z,\m) \equiv
\g_{F_i}\Big(\as(\m),\a(\m),\ln\frac{\z}{\m^2}\Big)
\nn\\
&\qquad\qquad=
- \g_i\big(\as(\m),\a(\m)\big) 
- \G_i\big(\as(\m),\a(\m)\big)\, \ln\frac{\z}{\m^2}
\,,
\nn\\
&\frac{d}{d\ln\z}\ln{\tilde F_i}(x,b_T;\z,\m) =
- D_i(L_\perp;\as(\m),\a(\m)) 
\,,
\nn\\
&\frac{d}{d\ln\m} D_i(L_\perp;\as(\m),\a(\m)) =
\G_i\big(\as(\m),\a(\m)\big)
\,,
\end{align}
where ${\tilde F_i}$ stands for any of the 32 leading-twist (un)polarized quark and gluon TMDPDFs or TMDFFs in coordinate space (or their derivatives, depending on their rank), which have the same evolution equations up to the flavor of the parton involved.
We notice that the full flavor universality of the evolution equations in pure QCD is broken when QED corrections are included, making them dependent on the flavor $i$ of the considered parton.
 
The cusp anomalous dimension $\G_i$ and the non-cusp piece $\g_i$ are known up to NNLO (${\cal O}(\as^3)$) in pure QCD \cite{Moch:2005id,Moch:2004pa} (numerical results for the cusp at NNNLO have recently been obtained in Ref.~\cite{Moch:2017uml}). 
The same holds for the $D_i$ term \cite{Echevarria:2015byo,Vladimirov:2016dll,Li:2016ctv}, which depends on $L_\perp=\ln(\m^2 b_T^2 e^{2\g_E}/4)$.
Notice that at small $b_T$ the $D_i$ term can be calculated perturbatively, but at large $b_T$ it has to be modeled and extracted from experimental data (see e.g. \cite{Echevarria:2014rua,Collins:2014jpa}).
We have not explicitly included any additional dependence on the resummation scales arising from QED corrections, but for simplicity set them equal to the corresponding ones in QCD ($\m$ and $\z$).

The two-dimensional evolution of the TMDs is performed in coordinate space as:
\begin{align}
{\tilde F_i}(x,b_T;\z,\m) &=
{\tilde F_i}(x,b_T;\z_0,\m_0)
\,
R_i(b_T;\z,\m,\z_0,\m_0)
\,,
\end{align}
where the evolution kernel $R_i$ is given by~\footnote{Here we choose a particular path $(\m_0,\z_0)\to(\m,\z)$. See \cite{Scimemi:2018xaf} for subtleties in this regard.}
\begin{align}
&R_i(b_T;\z,\m,\z_0,\m_0) 
\nn\\
&=
\exp\le\{
\int_{\m_0}^{\m} \frac{d\bar\m}{\bar\m}
\g_{F_i}\le(\as(\bar\m),\a(\bar\m),\ln\frac{\z}{\bar\m^2} \ri)\ri\}\,
\le(\frac{\z}{\z_0}\ri)^{-D_i(L_\perp;\as(\m_0),\a(\m_0))}
\,.
\end{align}

Once we have presented the evolution equations, our goal now is to provide the necessary QED corrections to $\g_{F_i}$ and $D_i$ in order to consistently perform the resummation at a given order.
In table~\ref{tab:resummation} we show the needed ingredients in pure QCD for the resummation at a given logarithmic accuracy.
The TMDs are customarily resummed in coordinate space by calculating them at the natural scale of the OPE, i.e. $\m_b\sim 1/b_T$, then evolved with the evolution kernel up to the relevant hard scale of the process, $Q$, and finally Fourier transformed back to momentum space.
This means that the lower scale $\m_b$ is integrated over a range that spans from approximately $\lqcd$ up to $Q$.

The relevance of the QED corrections depends on the relative size of $\a$ and $\as$, but given what we just explained, we cannot establish a quantitative relation between them that holds for all values of the running scale $\m_b$.
A conservative approach would be to consider $\a\sim\as^2$ at all scales.  
Another strategy would be to consider pure QED, mixed QCD$\times$QED and pure QCD contributions independently, without establishing any relation between $\a$ and $\as$ (see e.g. \cite{Cieri:2018sfk}).
In any case, we leave this for a future phenomenological study, and limit ourselves to providing the new perturbative ingredients which arise from the QED corrections.

\begin{table}[t!]
\begin{center}
\renewcommand{\arraystretch}{1.4}
\renewcommand{\tabcolsep}{0.2cm}
\begin{tabular}{c|c|c|c|c|c}\hline\hline
Order & $\G$ & $\g$ & $D$	 & $\tilde C_{i/j}$	& $\b$
\\
\hline
\hline
\hline
NLL & $\as^2$ & $\as^1$ & $\as^1$ & $\as^0$	& $\as^2$
\\
\hline
N$^2$LL & $\as^3$ & $\as^2$ & $\as^2$ & $\as^1$	& $\as^3$
\\
\hline
N$^3$LL & $\as^4$ & $\as^3$ & $\as^3$ & $\as^2$	& $\as^4$
\\
\hline\hline
\end{tabular}
\end{center}
\caption{\emph{Necessary perturbative ingredients to perform the resummation at a given logarithmic accuracy
}}
\label{tab:resummation}
\end{table}

Below we calculate the mixed QCD$\times$QED corrections to the anomalous dimension $\g_{F_i}$ (which includes the cusp $\G_i$ and the non-cusp $\g_i$) and the $D_i$ term at ${\cal O}(\as\a)$, and the pure QED ones up to ${\cal O}(\as^0\a^2)$.
This is achieved by profiting from the known results in pure QCD at ${\cal O}(\as^2\a^0)$.
From now on we will write the perturbative expansion of any function as 
$$A(\as,\a)=
\sum_{n,m} A^{(n,m)} \Big(\frac{\as}{4\pi}\Big)^n \Big(\frac{\a}{4\pi}\Big)^m
\,.
$$
In practice, we consider the relevant Feynman diagrams which contribute to each quantity at ${\cal O}(\as^2\a^0)$ and take the corresponding Abelian limit, by replacing one gluon by a photon, or two gluons by two photons.
Then we replace the needed color factors by the corresponding electromagnetic charge factors.
In \cite{deFlorian:2015ujt,deFlorian:2016gvk} this is referred to as an \emph{abelianization algorithm}.

\begin{figure*}[t!]
\centering
\subfloat[][]{
\includegraphics[width=0.2\textwidth]{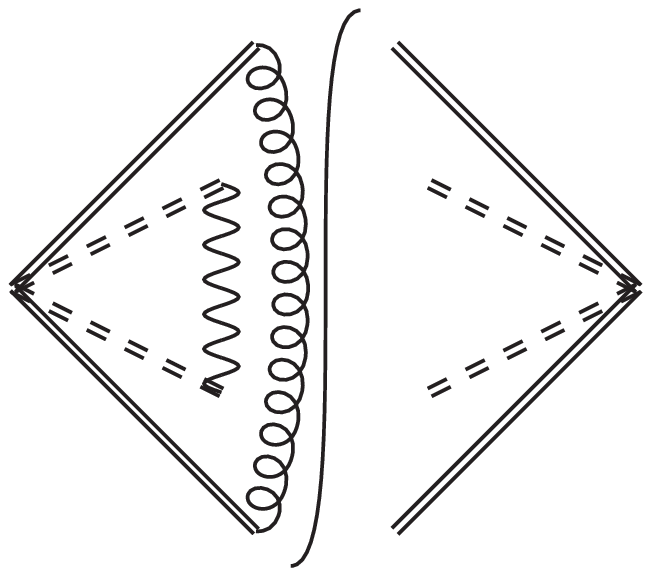}    
\label{fig:soft-a}}
\subfloat[][]{
\includegraphics[width=0.2\textwidth]{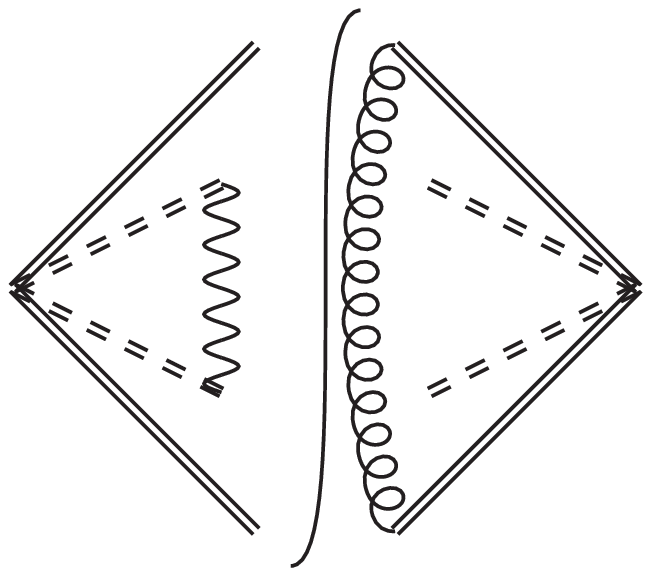}    
\label{fig:soft-b}}
\subfloat[][]{
\includegraphics[width=0.2\textwidth]{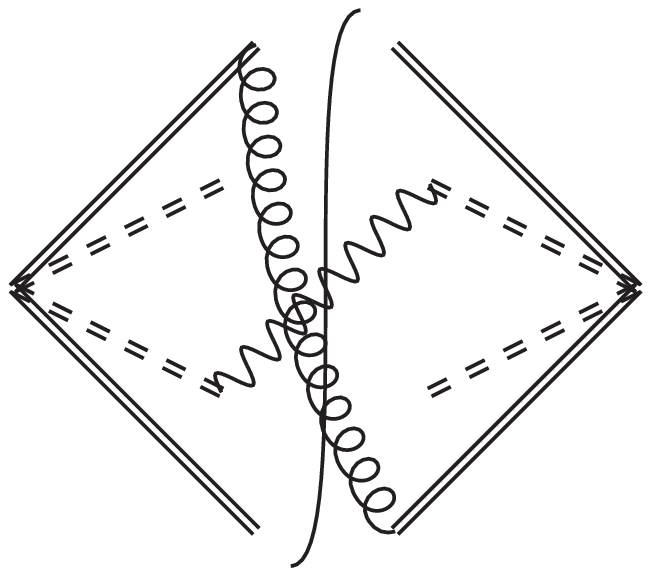}    
\label{fig:soft-b}}
\subfloat[][]{
\includegraphics[width=0.2\textwidth]{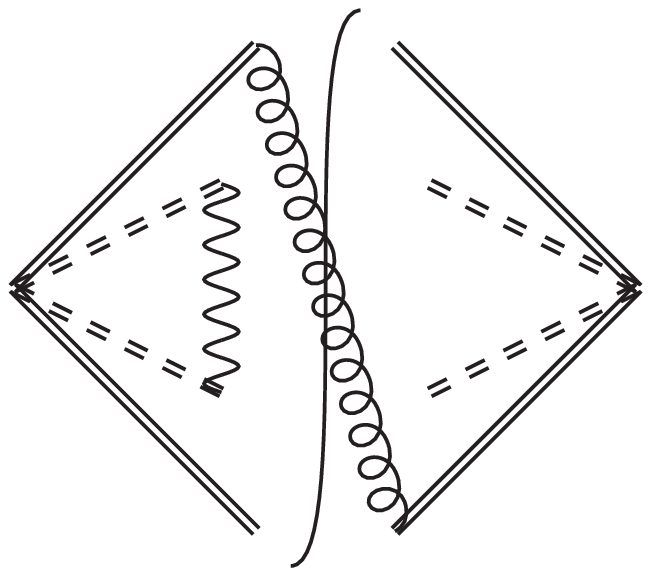}    
\label{fig:soft-d}}
\subfloat[][]{
\includegraphics[width=0.2\textwidth]{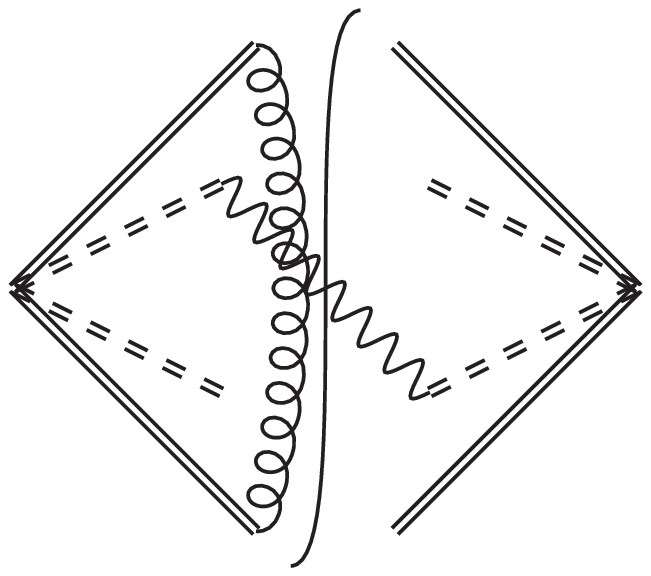}    
\label{fig:soft-e}}
\caption{\emph{Feynman diagrams for the soft function at ${\cal O}(\as\a)$.
Double solid lines stand for gluon Wilson lines, while double-dashed lines for photon Wilson lines.
Corresponding conjugated and crossed diagrams should be added.}}
\label{fig:soft}
\end{figure*}

Let us start with the $D_i$ term, which can be calculated only from the soft function \cite{Echevarria:2015usa}.
This is due to the fact that the $D_i$ term controls the evolution in the rapidity scale $\z$, which is a remnant of the arbitrary splitting of the soft function in rapidity space.

At leading order in $\a$, the $D_i$ term can be obtained from its analogous expression in pure QCD at order $\as$, since basically one needs to take the soft matrix element and calculate it at ${\cal O}(\as^0\a)$.
The result is (see Appendix):
\begin{align}
D_i^{(0,1)}(L_\perp) &=
\frac{\G_i^{(0,1)}}{2} L_\perp
\,,
\end{align}
where $\G_i^{(0,1)} = 4 Q_i^2$, with $Q_i$ the fractional electric charge of the considered $i$ quark.
We notice the flavor-dependence introduced by the QED corrections which, as already mentioned, breaks the flavor universality of the TMD evolution in pure QCD.

In order to obtain $D_i^{(1,1)}$, we realize that the soft function in perturbation theory in QCD$\times$QED can be written in a ``factorized'' form as
\begin{align}
\label{eq:SFfactorized}
S_i(k_{sT}) =&
\int\frac{d^2\vecb y_\perp}{(2\pi)^2}
e^{i\vecbe y _\perp \cdot \vecbe k _{s\perp}}
\tilde S^{QCD}(y_T)\,
\tilde S_i^{QED}(y_T)
+ {\cal O}(\as^n\a^m)\Big|_{n\cdot m>1}
\,,
\end{align}
where
\begin{align}
\tilde S^{QCD}(y_T) =&
\frac{{\rm Tr}_c}{N_c}
\sandwich{0}{\le[S_n^{T\dagger} S_\bn^T \ri]
(0^+,0^-,\vecb y_\perp)\le[S^{T\dagger}_\bn S_n^T \ri](0)}
{0}
\,,
\nn\\
\tilde S_i^{QED}(y_T) =&
\sandwich{0}{\le[
\widehat{S}_{i,n}^{T\dagger}
\widehat{S}_{i,\bn}^T \ri]
(0^+,0^-,\vecb y_\perp)\le[
\widehat{S}^{T\dagger}_{i,\bn}
\widehat{S}_{i,n}^T \ri](0)}
{0}
\,.
\end{align}
This factorized expression captures all contributions to the soft function at ${\cal O}(\as^n\a^0)$ and ${\cal O}(\as^0\a^n)$ for any $n$, and also the first mixed ones at ${\cal O}(\as\a)$.
Notice that the latter come solely from the multiplication of the two factorized matrix elements $\tilde S^{QCD}$ and $\tilde S_i^{QED}$, because there is no contribution at this order which can come from an interaction between gluon and photon Wilson lines, as shown by the representative Feynman diagrams in fig.~\ref{fig:soft}.
At higher mixed orders in $\as$ and $\a$ however, there will indeed be some of the contributions which cannot be casted in the factorized expression above.

On the other hand, the soft function has a special structure in perturbation theory.
In particular, using the $\d$-regulator, it can be written to all orders in perturbation theory as~\cite{Echevarria:2012js,Echevarria:2015byo}
\begin{align}
\ln \tilde S_i(b_T) &=
A_i(L_\perp;\as,\a) + 2D_i(L_\perp;\as,\a)\ln\frac{\d^+\d^-}{\m^2}
\,,
\end{align}
where we have included also the QED corrections, $A_i$ is a generic function and $D_i$ the already introduced function which controls the rapidity evolution of TMDs.
Combining this result with the expression in \eqref{eq:SFfactorized}, it is easy to see that
\begin{align}
D_i^{(1,1)}(L_\perp) &= 0
\,.
\end{align}

\begin{figure*}[t!]
\centering
\subfloat[][]{
\includegraphics[width=0.2\textwidth]{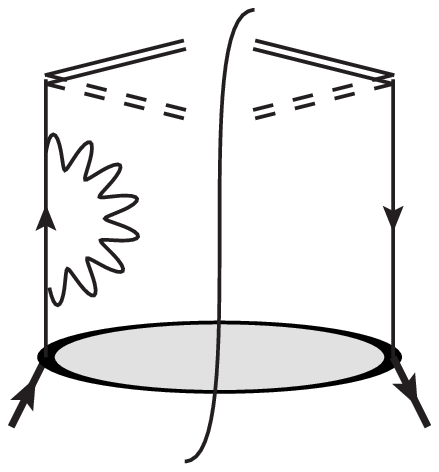}    
\label{fig:diagrams-a}}
\subfloat[][]{
\includegraphics[width=0.2\textwidth]{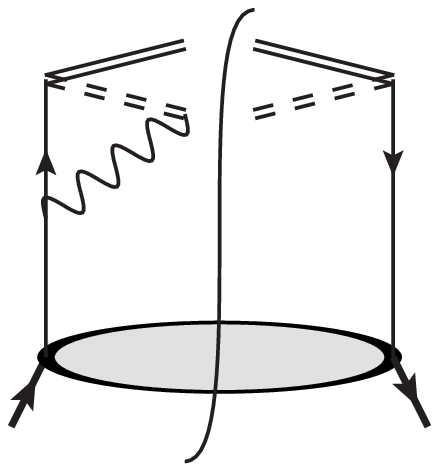}    
\label{fig:diagrams-b}}
\subfloat[][]{
\includegraphics[width=0.2\textwidth]{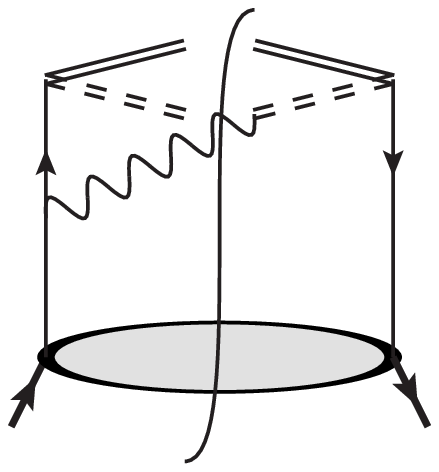}    
\label{fig:diagrams-c}}
\subfloat[][]{
\includegraphics[width=0.2\textwidth]{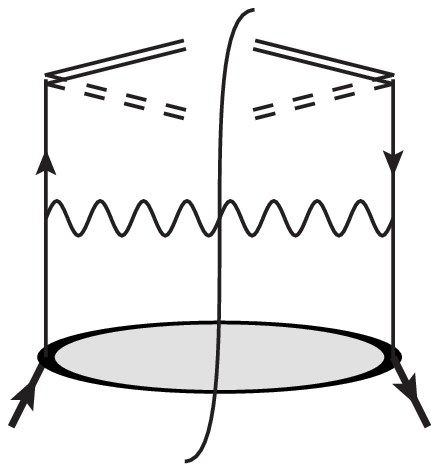}    
\label{fig:diagrams-d}}
\subfloat[][]{
\includegraphics[width=0.2\textwidth]{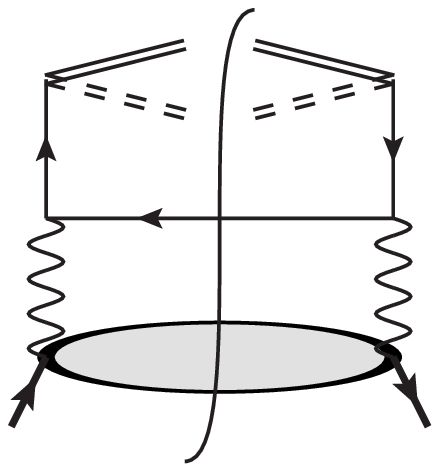}    
\label{fig:diagrams-e}}
\caption{\emph{Feynman diagrams for the TMDPDF at ${\cal O}(\as^0\a)$.
Double solid lines represent gluon Wilson lines, while double dashed lines stand for photon Wilson lines.}}
\label{fig:diagrams}
\end{figure*}

Finally, the $D_i$ term at ${\cal O}(\a^2)$ can again be obtained from its analogous expression in QCD.
The result is
\begin{align}
D_i^{(0,2)}(L_\perp) &=
\frac{\G_i^{(0,1)}}{4\widehat\b^{(0,1)}} \le(\widehat\b^{(0,1)} L_\perp \ri)^2
+ \le(\frac{\G_i^{(0,2)}}{2\widehat\b^{(0,1)}}\ri) \le(\widehat\b^{(0,1)} L_\perp \ri)
+ D_i^{(0,2)}(0)
\,,
\end{align}
where (see appendix for $D^{(2,0)}(0)$ and $\G^{(2,0)}$)
\begin{align}
D_i^{(0,2)}(0) &=
- \left(\frac{112}{27}\right)
Q_i^2\big[N_c \sum_j^{n_f} Q_j^2 + n_l Q_l^2\big]
\,, \nn\\
\G_i^{(0,2)}/\G_i^{(0,1)}&=
-\frac{20}{9} \big[N_c \sum_j^{n_f} Q_j^2 + n_l Q_l^2\big]
\,,
\end{align}
where $Q_{j}$ is the fractional charge of the active quarks ($n_f$) and $Q_{l}=1$ the one of the active leptons ($n_l$) in a fermion loop.
We need as well the QCD and QED $\b$ functions and the mixed contributions at ${\cal O}(\as\a)$:
\begin{align}
\frac{d\ln\as}{d\ln\m^2} &\equiv
\b\big(\as(\m),\a(\m)\big) =
- \sum_{n,m} \b^{(n,m)} \left( \frac{\as}{4\pi} \right)^{n} \left( \frac{\a}{4\pi} \right)^m
\,,\\
\frac{d\ln\a}{d\ln\m^2} &\equiv
\widehat\b\big(\as(\m),\a(\m)\big) =
- \sum_{n,m} \widehat\b^{(n,m)} \left( \frac{\as}{4\pi} \right)^{n} \left( \frac{\a}{4\pi} \right)^{m}
\,,
\end{align}
with the coefficients (see e.g. \cite{Kilgore:2011pa,Kilgore:2013uta})
\begin{align}
\b^{(1,1)} &=
-2\sum_j^{n_f} Q_j^2
\,,\\
\widehat\b^{(0,1)} &=
-\frac{4}{3} \big[N_c \sum_j^{n_f} Q_j^2 + n_l Q_l^2\big]
\,,\\
\widehat\b^{(0,2)} &=
-4 \big[N_c \sum_j^{n_f} Q_j^4 + n_l Q_l^4\big]
\,,\\
\widehat\b^{(1,1)} &=
-4 C_F N_c \sum_j^{n_f} Q_j^2
\,.
\end{align}

We turn now to the QED corrections to the anomalous dimension $\g_{F_i}$.
At leading order in $\a$, the anomalous dimension is calculated from the virtual diagrams in figs.~\ref{fig:diagrams-a}-\ref{fig:diagrams-b}, together with the corresponding one-loop diagram of the soft function in pure QED, and apart from charge factors, it has an analogous expression to that in QCD.
The result is:
\begin{align}
\g_i^{(0,1)} &= -6Q_i^2
\,,
\nn\\
\G_i^{(0,1)} &= 4Q_i^2
\,.
\end{align}

\begin{figure*}[t!]
\centering
\subfloat[][]{
\includegraphics[width=0.2\textwidth]{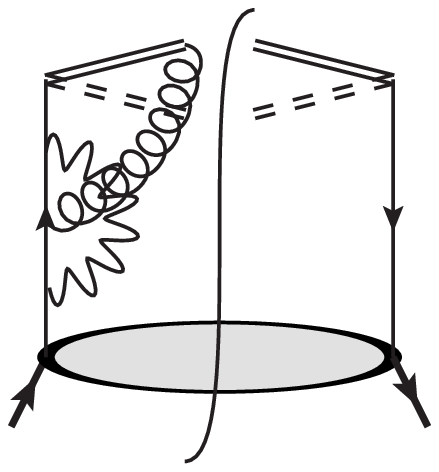}    
\label{fig:ad11-a}}
\subfloat[][]{
\includegraphics[width=0.2\textwidth]{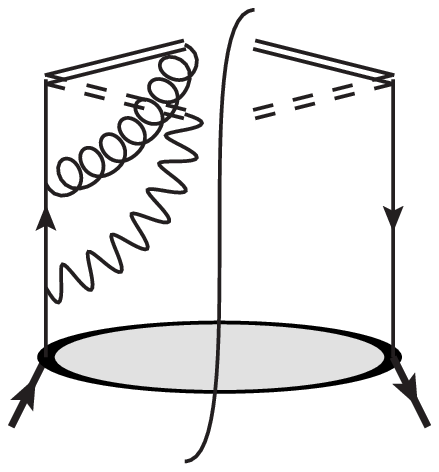}    
\label{fig:ad11-b}}
\subfloat[][]{
\includegraphics[width=0.2\textwidth]{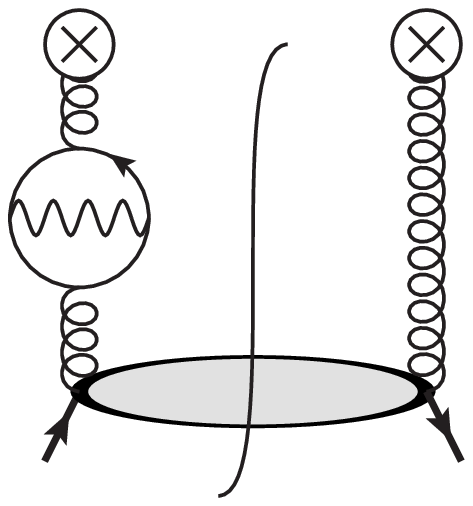}    
\label{fig:ad11g}}
\caption{\emph{(a)-(b): representative Feynman diagrams which contribute to $\g_{F_i}$ for quark TMDs at ${\cal O}(\as\a)$} through $J_{i/P}$. (c): similar but for $\g_{F_g}$ for gluon TMDs, where the crossed vertex stands for the collinear gluon Wilson line in $J_{g/P}$.}
\label{fig:ad11}
\end{figure*}

At ${\cal O}(\as\a)$ one can take the results $\g_i^{(0,2)}$ and $\G_i^{(0,2)}$ from the appendix and apply the recipe to translate QCD color factors to QED charge factors.
In this case, one needs to replace one gluon by one photon in the two-gluon diagrams that contribute in pure QCD.
Proceeding in this way we obtain
\begin{align}
\g_i^{(1,1)} &=
2 C_F Q_i^2\(-3+4\pi^2-48\zeta_3\)
\,,
\nn\\
\G_i^{(1,1)} &= 0
\,.
\end{align}
In order to obtain $\g_i^{(1,1)}$, we have noted that by color/charge conservation there cannot be fermion-loop diagrams at ${\cal O}(\as\a)$, nor in the collinear matrix element $J_{i/P}$ nor in the soft function (see fig.~\ref{fig:soft}), so the term proportional to $T_F n_f$ in $\g_i^{(0,2)}$ (see Appendix) does not contribute after taking the (partial) Abelian limit.
In fact, only the diagrams in figs.~\ref{fig:ad11-a} and~\ref{fig:ad11-b} (and similar ones) and the ones in fig.~\ref{fig:soft} contribute.
Also, we added an additional factor of 2 to $\g_i^{(1,1)}$ as compared to $\g_i^{(2,0)}$, since there are two ways of replacing the 2 internal gluons in the relevant Feynman diagrams by a gluon and a photon. 
We note that in \cite{Kilgore:2011pa,Kilgore:2013uta} it is also found that $\G_i^{(1,1)}=0$.

Finally, we provide the two-loop QED corrections to the anomalous dimension $\g_{F_i}$.
Again, they can be obtained from the expressions in pure QCD $\g_i^{(2,0)}$ and $\G_i^{(2,0)}$, by replacing in the relevant Feynman diagrams the two gluons by two photons.
The results are:
\begin{align}
\g_i^{(0,2)} &=
Q_i^4\(-3+4\pi^2-48\zeta_3\)
\nn\\
&+Q_i^2 \big[N_c \sum_j^{n_f} Q_j^2 + n_l Q_l^2\big] \(\frac{260}{27}+\frac{4\pi^2}{3}\)
\,,
\nn\\
\G_i^{(0,2)}/\G_i^{(0,1)}&=
-\frac{20}{9} \big[N_c \sum_j^{n_f} Q_j^2 + n_l Q_l^2\big]
\,.
\end{align}

Now we turn our attention to the evolution kernel for gluon TMDs \cite{Echevarria:2015uaa}, which is analogous to the one of quark TMDs shown at the beginning of this section in \eqref{eq:quarkTMDevo} with $i=g$.
The QED corrections to the relevant anomalous dimension $\g_{F_g}$ and the $D_g$ term only appear, however, at ${\cal O}(\as\a)$ and beyond, and there are no pure QED corrections.
This is because, given color/charge conservation, one needs in the relevant Feynman diagrams at least one closed quark loop, where a gluon inside can be replaced by a photon.
In fact, only the collinear matrix element $J_{g/P}$ contributes at ${\cal O}(\as\a)$, see fig.~\ref{fig:ad11g}, since the soft function for gluon TMDs starts to contribute at ${\cal O}(\as^2\a)$.

The non-cusp anomalous dimension for gluon TMDs in pure QCD at ${\cal O}(\as^2)$ is (see e.g. \cite{Echevarria:2015uaa})
\begin{align}
\g_g^{(2,0)} &= 
2C_A^2 \left( -\frac{692}{27} + \frac{11\pi^2}{18}
+ 2\zeta_3 \right) 
+ 2C_A T_F n_f \left( \frac{256}{27} - \frac{2\pi^2}{9} \right)
\nn\\
&+ 8 C_F T_F n_f
\,,
\end{align}
from which we obtain
\begin{align}
\g_g^{(1,1)} &= 
8 T_F \sum_j^{n_f} Q_j^2
\,.
\end{align}
Notice that there is no factor $N_c$, because we are replacing an inner gluon by a photon in a closed fermion loop inside a gluon line, which remains.
Thus we do not need to account for the color multiplicity, as in the case of a fermion loop inside a photon  line.

The cusp anomalous dimension for gluons does not receive QED corrections at ${\cal O}(\as\a)$: $\G_g^{(1,1)}=0$.
Neither does the $D^g$ term, since the soft function (for gluon TMDs) from which it can be obtained is zero at this order, and thus $D_g^{(1,1)}=0$.

\section{QED corrections for $f_1$ at large transverse momentum}
\label{sec:tmdpdf}

When the transverse momentum is large, i.e. the impact parameter is small, the unpolarized quark TMDPDF can be expanded through an operator product expansion (OPE) in terms of its corresponding integrated counterparts, the collinear quark/gluon PDFs:
\begin{align}\label{eq:ope}
{\tilde f}_{i/A}(x,b_T;\z,\m) &=
\sum_{j=q,\bar q, g, \g}
{\tilde C}_{i/j}(x,b_T;\z,\m)\otimes
f_{j/A}(x;\m)\,
\nn\\
&+ {\cal O}\Big(b_T\lqcd\Big)
\,,
\end{align}
where $\z_0\sim \m_0^2 \sim \m_b^2$, with $\m_b=2e^{-\g_E}/b_T$ the natural scale at which the OPE is performed.
Notice that we have included the contribution of the photon PDF in the OPE.
The coefficients $\tilde C_{i/j}$ (excluding $j=\g$, which is part of the aim of this paper) are currently known in pure QCD up to ${\cal O}(\as^2)$ (see e.g. \cite{Echevarria:2016scs}).

We consider the inclusion of the QED corrections at LO, which will modify the evolution/resummation of the TMDPDF.
As already discussed, they will contribute to the evolution of the TMDs.
In addition, they will affect the matching of the TMDPDF onto its collinear counterparts.

The photon PDF will have to be included in the sum over partons of the OPE, as introduced above.
This will amount to have a matching coefficient of the quark TMDPDF onto the photon PDF, which is:
\begin{align}
{\tilde C}_{i/\g}^{(0,1)}(x,b_T;\m) &=
N_c Q_i^2\Big[
-2L_T\big(x^2+(1-x)^2\big) + 4x(1-x)
\Big]
\,.
\end{align}
This coefficient can be calculated directly by an explicit perturbative calculation of both the quark TMDPDF and the photon PDF (by subtraction using \eqref{eq:ope}), or obtained for instance from ${\tilde C}_{i/g}^{(1,0)}$ in eq.~(39) in \cite{Becher:2010tm} or eq.~(A10) in \cite{Aybat:2011zv} by replacing the color factor ($T_F$) by the proper charge factor ($N_c Q_i^2$).
Notice that we have included a factor $N_c$, which accounts for the color multiplicity of the quark~\footnote{When calculating the QCD version of the diagram in fig.~\ref{fig:diagrams-e}, with the photons replaced by gluons, one sums over the color of quarks and averages over the color of gluons, while in fig.~\ref{fig:diagrams-e} there is no average to be performed, but the sum over the color of quarks still gives the factor $N_c$ mentioned.}.

Also the matching coefficient of the TMDPDF onto the quark PDF will have to be modified, which at ${\cal O}(\a)$ is analogous to the pure QCD one ${\tilde C}_{i/i}^{(1,0)}$.
In fact, one has:
\begin{align}
{\tilde C}_{i/i}^{(0,1)}(x,b_T;\m) &=
Q_i^2\Bigg[ 
\d(1-x)\le( - L_T^2 + 3L_T + 2L_T\ln\frac{\m^2}{Q^2} - \frac{\pi^2}{6}\ri)
\nn\\
&- 2L_T {\cal P}_{i\leftarrow i} + 2(1-x)
\Bigg]
\,.
\end{align}
This can also be directly calculated by considering the quark TMDPDF and the quark PDF up to ${\cal O}(\a)$.
The inclusion of the newly derived coefficients ${\tilde C}_{i/i}^{(0,1)}$ and ${\tilde C}_{i/\g}^{(0,1)}$ will be necessary at N$^3$LL and beyond, if one takes the recipe in table~\ref{tab:resummation} with the counting $\a\sim\as^2$.
The QED corrections to the OPE coefficients at ${\cal O}(\as\a)$, even if possible to derive from the already known ones at ${\cal O}(\as^2)$ in pure QCD, will only enter at N$^4$LL, which is far beyond the current achievable theoretical precision, and even more of the experimental one.

We end this section by noting that the leading-order QED corrections to the OPE coefficients of the helicity and transversity TMDPDFs and the unpolarized, helicity and transversity TMDFFs can be calculated similarly, starting from the known expressions in pure QCD.
Also the ones corresponding to polarized TMDs, as the Sivers or Collins functions.
We leave this for a future effort.

\section{Conclusions}
\label{sec:conclusions}

By extending the operator definition of TMDs from pure QCD to QCD$\times$QED, we have calculated the mixed corrections to the relevant anomalous dimensions at ${\cal O}(\as\a)$ and the pure QED ones up to ${\cal O}(\a^2)$.
They apply to all the leading-twist unpolarized and polarized quark and gluon TMD parton distribution and fragmentation functions.
To do so, we have taken advantage of the known results in pure QCD at ${\cal O}(\as^2)$ and replaced, in the relevant Feynman diagrams, one internal gluon by a photon, or two gluons by two photons, and then recalculated the proper color/charge factors.
These corrections depend on the flavor of the considered parton, and thus break the full flavor universality of the TMD evolution kernel in pure QCD (but not the spin universality).

In addition, we have also calculated the leading-order QED corrections at ${\cal O}(\a)$ to the matching coefficients of the unpolarized quark TMDPDF onto its integrated counterparts, again profiting from known results in pure QCD.
Following the same procedure, the analogous QED corrections to other TMDs can be calculated in a straightforward way.

The newly calculated corrections will make it possible to perform very precise resummations of large logarithms for any of the leading-twist TMDs, in view of the precision needs of future planned experiments~\cite{Accardi:2012qut,Dudek:2012vr,Brodsky:2012vg,Kikola:2017hnp,Hadjidakis:2018ifr}.

Finally, we note that the new results apply as well to the evolution of generalized TMDs, since their evolution is analogous to the one of the TMDs~\cite{Echevarria:2016mrc}.

{\it \textbf{Acknowledgements.}}
We thank Joan Soto for his contributions to the initial stage of this project.
AB and MGE are supported by the European Research Council (ERC) under the European Union's Horizon 2020 research and innovation program (grant agreement No. 647981, 3DSPIN).
MGE is supported by the Marie Sk\l odowska-Curie grant \emph{GlueCore} (grant agreement No. 793896).

\appendix
\section{Anomalous dimensions}
\label{app}

The cusp anomalous dimension $\G$ is:
\begin{align}
\Gamma^{(1,0)}&=4C_F
\,,
\nn\\ 
\Gamma^{(2,0)}/\Gamma^{(1,0)}&=
\(\frac{67}{9}-\frac{\pi^2}{3}\)C_A-\frac{20}{9}T_F n_f
\,,
\nn\\
\G^{(3,0)}/\Gamma^{(1,0)} &=
 C_A^2 \left( \frac{245}{6} - \frac{134\pi^2}{27}
+ \frac{11\pi^4}{45} + \frac{22}{3}\,\zeta_3 \right)
\nn\\
&+ C_A T_F n_f  \left( - \frac{418}{27} + \frac{40\pi^2}{27}
- \frac{56}{3}\,\zeta_3 \right)
\nn\\
&
+ C_F T_F n_f \left( - \frac{55}{3} + 16\zeta_3 \right)
- \frac{16}{27}\,T_F^2 n_f^2
\,.
\end{align}

The non-cusp anomalous dimension $\g$:
\begin{align}
\gamma^{(1,0)}&=-6C_F,
\nn\\
\gamma^{(2,0)}&=
C_F^2\(-3+4\pi^2-48\zeta_3\)+C_FC_A\(-\frac{961}{27}-\frac{11\pi^2}{3}+52\zeta_3\)
\nn\\
&
+C_FT_Fn_f\(\frac{260}{27}+\frac{4\pi^2}{3}\),
\\
\gamma^{(3,0)}&=
C_F^3 \left( -29 - 6\pi^2 - \frac{16\pi^4}{5}
- 136\zeta_3 + \frac{32\pi^2}{3}\,\zeta_3 + 480\zeta_5 \right)
\nn\\
&
+ C_F^2 C_A \Bigg( - \frac{151}{2} + \frac{410\pi^2}{9}
+ \frac{494\pi^4}{135} - \frac{1688}{3}\,\zeta_3
\nn\\
&
- \frac{16\pi^2}{3}\,\zeta_3 - 240\zeta_5 \Bigg)
\nn\\
&
+ C_F C_A^2 \Bigg( - \frac{139345}{1458} - \frac{7163\pi^2}{243}
- \frac{83\pi^4}{45} + \frac{7052}{9}\,\zeta_3
\nn\\
&
- \frac{88\pi^2}{9}\,\zeta_3 - 272\zeta_5 \Bigg)
\nn\\
&
+ C_F^2 T_F n_f \left( \frac{5906}{27} - \frac{52\pi^2}{9}
- \frac{56\pi^4}{27} + \frac{1024}{9}\,\zeta_3 \right)
\nn\\
&
+ C_F C_A T_F n_f \left( - \frac{34636}{729}
+ \frac{5188\pi^2}{243} + \frac{44\pi^4}{45} - \frac{3856}{27}\,\zeta_3
\right)
\nn\\
&
+ C_F T_F^2 n_f^2 \left( \frac{19336}{729} - \frac{80\pi^2}{27}
- \frac{64}{27}\,\zeta_3 \right)
\,.
\end{align}

The coefficients of the $D$ term are
\begin{align}\label{eq:dcoeffs}
D^{(1,0)}(L_\perp) &=
\frac{\G^{(1,0)}}{2 \b^{(1,0)}} \le(\b^{(1,0)} L_\perp \ri)
+ D^{(1,0)}(0)
\,,
\nn\\
D^{(2,0)}(L_\perp) &=
\frac{\G^{(1,0)}}{4\b^{(1,0)}} \le(\b^{(1,0)} L_\perp \ri)^2
+ \le(\frac{\G^{(2,0)}}{2\b^{(1,0)}} + D^{(1,0)}(0)\ri) \big(\b^{(1,0)} L_\perp \big)
\nn\\
&
+ D^{(2,0)}(0)
\,,
\nn\\
D^{(3,0)}(L_\perp) &=
\frac{\G^{(1,0)}}{6\b^{(1,0)}} \le(\b^{(1,0)} L_\perp \ri)^3
\nn\\
&
+ \frac{1}{2}\le(\frac{\G^{(2,0)}}{\b^{(1,0)}} + \frac{1}{2}\frac{\G^{(1,0)}\b^{(2,0)}}{[\b^{(1,0)}]^2} + 2D^{(1,0)}(0) \ri)
\le(\b^{(1,0)} L_\perp \ri)^2
\nn\\
&
+
\frac{1}{2} \le( 4D^{(2,0)}(0) + \frac{\b^{(2,0)}}{\b^{(1,0)}} 2D^{(1,0)}(0)
+ \frac{\G^{(3,0)}}{\b^{(1,0)}}\ri) \le(\b^{(1,0)} L_\perp \ri)
\nn\\
&
+ D^{(3,0)}(0)
\,,
\end{align}
with $L_\perp=\ln\frac{\m^2 b^2}{4e^{-2\g_E}}$ and
\begin{align}
D^{(1,0)}(0) &= 0
\,,
\nn\\
D^{(2,0)}(0) &=
C_F C_A \left(\frac{404}{27}-14\z_3\right)
- \left(\frac{112}{27}\right)C_F T_F n_f
\,,
\nn\\
D^{(3,0)}(0)&= 
\frac{-1}{2}C_FC_A^2 \Bigg(-\frac{176}{3}\zeta_3\zeta_2+\frac{6392\zeta_2}{81}+\frac{12328\zeta_3}{27}
\nn\\
&
+\frac{154\zeta_4}{3}-192 \zeta_5-\frac{297029}{729}\Bigg)
\nn \\ 
&
- C_FC_A T_F n_f\left(-\frac{824\zeta_2}{81}-\frac{904\zeta_3}{27}+\frac{20\zeta_4}{3}+\frac{62626}{729}\right)
\nn\\
&
- 2 T_F^2 n_f^2 \left(-\frac{32\zeta_3}{9}-\frac{1856}{729}\right)
\nn \\ 
&
- C_F^2 T_F n_f \left(\frac{-304\zeta_3}{9}-16\zeta_4+\frac{1711}{27}\right)
\,.
\end{align}
The result for $D^{(3,0)}(0)$ has been recently computed in \cite{Li:2016ctv}. 
The rest can be found also in \cite{Echevarria:2012pw}.

Finally, the coefficients for the QCD $\beta$-function are 
\begin{align}
\b^{(1,0)} &=
\frac{11}{3}\,C_A - \frac43\,T_F n_f
\,,
\nn\\
\b^{(2,0)} &=
\frac{34}{3}\,C_A^2 - \frac{20}{3}\,C_A T_F n_f
- 4 C_F T_F n_f
\,,
\nn\\
\b^{(3,0)} &=
\frac{2857}{54}\,C_A^3 + \left( 2 C_F^2
- \frac{205}{9}\,C_F C_A - \frac{1415}{27}\,C_A^2 \right) T_F n_f
\nn\\
&
+ \left( \frac{44}{9}\,C_F + \frac{158}{27}\,C_A \right) T_F^2 n_f^2
\,,
\nn\\
\b^{(4,0)} &=
\frac{149753}{6} + 3564\zeta_3
- \left( \frac{1078361}{162} + \frac{6508}{27}\,\zeta_3 \right) n_f
\nn\\
&
+ \left( \frac{50065}{162} + \frac{6472}{81}\,\zeta_3 \right) n_f^2
+ \frac{1093}{729}\,n_f^3
\,,
\end{align}
where for $\b^{(4,0)}$ we have used $N_c=3$ and $T_F=\frac{1}{2}$.

\bibliographystyle{utphys}  
\bibliography{references}

\providecommand{\href}[2]{#2}\begingroup\raggedright\begin{thebibliography}{10}

\bibitem{Angeles-Martinez:2015sea}
R.~Angeles-Martinez {\em et~al.}, ``{Transverse Momentum Dependent (TMD) parton
  distribution functions: status and prospects},''
  \href{http://dx.doi.org/10.5506/APhysPolB.46.2501}{{\em Acta Phys. Polon.}
  {\bfseries B46} no.~12, (2015) 2501--2534},
\href{http://arxiv.org/abs/1507.05267}{{\ttfamily arXiv:1507.05267 [hep-ph]}}.

\bibitem{Rogers:2015sqa}
T.~C. Rogers, ``{An overview of transverse-momentum-dependent factorization and
  evolution},'' \href{http://dx.doi.org/10.1140/epja/i2016-16153-7}{{\em Eur.
  Phys. J.} {\bfseries A52} no.~6, (2016) 153},
\href{http://arxiv.org/abs/1509.04766}{{\ttfamily arXiv:1509.04766 [hep-ph]}}.

\bibitem{Diehl:2015uka}
M.~Diehl, ``{Introduction to GPDs and TMDs},''
  \href{http://dx.doi.org/10.1140/epja/i2016-16149-3}{{\em Eur. Phys. J.}
  {\bfseries A52} no.~6, (2016) 149},
\href{http://arxiv.org/abs/1512.01328}{{\ttfamily arXiv:1512.01328 [hep-ph]}}.

\bibitem{Ji:2004wu}
X.-d. Ji, J.-p. Ma, and F.~Yuan, ``{QCD factorization for semi-inclusive
  deep-inelastic scattering at low transverse momentum},''
  \href{http://dx.doi.org/10.1103/PhysRevD.71.034005}{{\em Phys. Rev.}
  {\bfseries D71} (2005) 034005},
\href{http://arxiv.org/abs/hep-ph/0404183}{{\ttfamily arXiv:hep-ph/0404183
  [hep-ph]}}.

\bibitem{Collins:2007nk}
J.~Collins and J.-W. Qiu, ``{$k_{T}$ factorization is violated in production of
  high-transverse-momentum particles in hadron-hadron collisions},''
  \href{http://dx.doi.org/10.1103/PhysRevD.75.114014}{{\em Phys. Rev.}
  {\bfseries D75} (2007) 114014},
\href{http://arxiv.org/abs/0705.2141}{{\ttfamily arXiv:0705.2141 [hep-ph]}}.

\bibitem{Rogers:2010dm}
T.~C. Rogers and P.~J. Mulders, ``{No Generalized TMD-Factorization in
  Hadro-Production of High Transverse Momentum Hadrons},''
  \href{http://dx.doi.org/10.1103/PhysRevD.81.094006}{{\em Phys. Rev.}
  {\bfseries D81} (2010) 094006},
\href{http://arxiv.org/abs/1001.2977}{{\ttfamily arXiv:1001.2977 [hep-ph]}}.

\bibitem{Becher:2010tm}
T.~Becher and M.~Neubert, ``{{Drell-Yan} Production at Small $q_T$, Transverse
  Parton Distributions and the Collinear Anomaly},''
  \href{http://dx.doi.org/10.1140/epjc/s10052-011-1665-7}{{\em Eur. Phys. J.}
  {\bfseries C71} (2011) 1665},
\href{http://arxiv.org/abs/1007.4005}{{\ttfamily arXiv:1007.4005 [hep-ph]}}.

\bibitem{Chiu:2012ir}
J.-Y. Chiu, A.~Jain, D.~Neill, and I.~Z. Rothstein, ``{A Formalism for the
  Systematic Treatment of Rapidity Logarithms in Quantum Field Theory},''
  \href{http://dx.doi.org/10.1007/JHEP05(2012)084}{{\em JHEP} {\bfseries 05}
  (2012) 084},
\href{http://arxiv.org/abs/1202.0814}{{\ttfamily arXiv:1202.0814 [hep-ph]}}.

\bibitem{Mantry:2010bi}
S.~Mantry and F.~Petriello, ``{Transverse Momentum Distributions in the
  Non-Perturbative Region},''
  \href{http://dx.doi.org/10.1103/PhysRevD.84.014030}{{\em Phys. Rev.}
  {\bfseries D84} (2011) 014030},
\href{http://arxiv.org/abs/1011.0757}{{\ttfamily arXiv:1011.0757 [hep-ph]}}.

\bibitem{Collins:2011zzd}
J.~Collins, ``{Foundations of perturbative QCD},''
{\em Camb. Monogr. Part. Phys. Nucl. Phys. Cosmol.} {\bfseries 32} (2011)
  1--624.

\bibitem{Aybat:2011zv}
S.~M. Aybat and T.~C. Rogers, ``{TMD Parton Distribution and Fragmentation
  Functions with QCD Evolution},''
  \href{http://dx.doi.org/10.1103/PhysRevD.83.114042}{{\em Phys. Rev.}
  {\bfseries D83} (2011) 114042},
\href{http://arxiv.org/abs/1101.5057}{{\ttfamily arXiv:1101.5057 [hep-ph]}}.

\bibitem{GarciaEchevarria:2011rb}
M.~G. Echevarria, A.~Idilbi, and I.~Scimemi, ``{Factorization Theorem For
  Drell-Yan At Low $q_T$ And Transverse Momentum Distributions
  On-The-Light-Cone},'' \href{http://dx.doi.org/10.1007/JHEP07(2012)002}{{\em
  JHEP} {\bfseries 07} (2012) 002},
\href{http://arxiv.org/abs/1111.4996}{{\ttfamily arXiv:1111.4996 [hep-ph]}}.

\bibitem{Echevarria:2012js}
M.~G. Echevarria, A.~Idilbi, and I.~Scimemi, ``{Soft and Collinear
  Factorization and Transverse Momentum Dependent Parton Distribution
  Functions},'' \href{http://dx.doi.org/10.1016/j.physletb.2013.09.003}{{\em
  Phys. Lett.} {\bfseries B726} (2013) 795--801},
\href{http://arxiv.org/abs/1211.1947}{{\ttfamily arXiv:1211.1947 [hep-ph]}}.

\bibitem{Vladimirov:2017ksc}
A.~Vladimirov, ``{Structure of rapidity divergences in multi-parton scattering
  soft factors},'' \href{http://dx.doi.org/10.1007/JHEP04(2018)045}{{\em JHEP}
  {\bfseries 04} (2018) 045},
\href{http://arxiv.org/abs/1707.07606}{{\ttfamily arXiv:1707.07606 [hep-ph]}}.

\bibitem{Echevarria:2012pw}
M.~G. Echevarria, A.~Idilbi, A.~Schaefer, and I.~Scimemi, ``{Model-Independent
  Evolution of Transverse Momentum Dependent Distribution Functions (TMDs) at
  NNLL},'' \href{http://dx.doi.org/10.1140/epjc/s10052-013-2636-y}{{\em Eur.
  Phys. J.} {\bfseries C73} no.~12, (2013) 2636},
\href{http://arxiv.org/abs/1208.1281}{{\ttfamily arXiv:1208.1281 [hep-ph]}}.

\bibitem{Echevarria:2014rua}
M.~G. Echevarria, A.~Idilbi, and I.~Scimemi, ``{Unified treatment of the QCD
  evolution of all (un-)polarized transverse momentum dependent functions:
  Collins function as a study case},''
  \href{http://dx.doi.org/10.1103/PhysRevD.90.014003}{{\em Phys. Rev.}
  {\bfseries D90} no.~1, (2014) 014003},
\href{http://arxiv.org/abs/1402.0869}{{\ttfamily arXiv:1402.0869 [hep-ph]}}.

\bibitem{Echevarria:2015byo}
M.~G. Echevarria, I.~Scimemi, and A.~Vladimirov, ``{Universal transverse
  momentum dependent soft function at NNLO},''
  \href{http://dx.doi.org/10.1103/PhysRevD.93.054004}{{\em Phys. Rev.}
  {\bfseries D93} no.~5, (2016) 054004},
\href{http://arxiv.org/abs/1511.05590}{{\ttfamily arXiv:1511.05590 [hep-ph]}}.

\bibitem{Li:2016ctv}
Y.~Li and H.~X. Zhu, ``{Bootstrapping Rapidity Anomalous Dimensions for
  Transverse-Momentum Resummation},''
  \href{http://dx.doi.org/10.1103/PhysRevLett.118.022004}{{\em Phys. Rev.
  Lett.} {\bfseries 118} no.~2, (2017) 022004},
\href{http://arxiv.org/abs/1604.01404}{{\ttfamily arXiv:1604.01404 [hep-ph]}}.

\bibitem{Vladimirov:2016dll}
A.~A. Vladimirov, ``{Correspondence between Soft and Rapidity Anomalous
  Dimensions},'' \href{http://dx.doi.org/10.1103/PhysRevLett.118.062001}{{\em
  Phys. Rev. Lett.} {\bfseries 118} no.~6, (2017) 062001},
\href{http://arxiv.org/abs/1610.05791}{{\ttfamily arXiv:1610.05791 [hep-ph]}}.

\bibitem{Catani:2013tia}
S.~Catani, L.~Cieri, D.~de~Florian, G.~Ferrera, and M.~Grazzini,
  ``{Universality of transverse-momentum resummation and hard factors at the
  NNLO},'' \href{http://dx.doi.org/10.1016/j.nuclphysb.2014.02.011}{{\em Nucl.
  Phys.} {\bfseries B881} (2014) 414--443},
\href{http://arxiv.org/abs/1311.1654}{{\ttfamily arXiv:1311.1654 [hep-ph]}}.

\bibitem{Gehrmann:2014yya}
T.~Gehrmann, T.~Luebbert, and L.~L. Yang, ``{Calculation of the transverse
  parton distribution functions at next-to-next-to-leading order},''
  \href{http://dx.doi.org/10.1007/JHEP06(2014)155}{{\em JHEP} {\bfseries 06}
  (2014) 155},
\href{http://arxiv.org/abs/1403.6451}{{\ttfamily arXiv:1403.6451 [hep-ph]}}.

\bibitem{Echevarria:2016scs}
M.~G. Echevarria, I.~Scimemi, and A.~Vladimirov, ``{Unpolarized Transverse
  Momentum Dependent Parton Distribution and Fragmentation Functions at
  next-to-next-to-leading order},''
  \href{http://dx.doi.org/10.1007/JHEP09(2016)004}{{\em JHEP} {\bfseries 09}
  (2016) 004},
\href{http://arxiv.org/abs/1604.07869}{{\ttfamily arXiv:1604.07869 [hep-ph]}}.

\bibitem{Bacchetta:2013pqa}
A.~Bacchetta and A.~Prokudin, ``{Evolution of the helicity and transversity
  Transverse-Momentum-Dependent parton distributions},''
  \href{http://dx.doi.org/10.1016/j.nuclphysb.2013.07.013}{{\em Nucl. Phys.}
  {\bfseries B875} (2013) 536--551},
\href{http://arxiv.org/abs/1303.2129}{{\ttfamily arXiv:1303.2129 [hep-ph]}}.

\bibitem{Gutierrez-Reyes:2017glx}
D.~Gutiérrez-Reyes, I.~Scimemi, and A.~A. Vladimirov, ``{Twist-2 matching of
  transverse momentum dependent distributions},''
  \href{http://dx.doi.org/10.1016/j.physletb.2017.03.031}{{\em Phys. Lett.}
  {\bfseries B769} (2017) 84--89},
\href{http://arxiv.org/abs/1702.06558}{{\ttfamily arXiv:1702.06558 [hep-ph]}}.

\bibitem{Buffing:2017mqm}
M.~G.~A. Buffing, M.~Diehl, and T.~Kasemets, ``{Transverse momentum in double
  parton scattering: factorisation, evolution and matching},''
  \href{http://dx.doi.org/10.1007/JHEP01(2018)044}{{\em JHEP} {\bfseries 01}
  (2018) 044},
\href{http://arxiv.org/abs/1708.03528}{{\ttfamily arXiv:1708.03528 [hep-ph]}}.

\bibitem{Gutierrez-Reyes:2018iod}
D.~Gutierrez-Reyes, I.~Scimemi, and A.~Vladimirov, ``{Transverse momentum
  dependent transversely polarized distributions at
  next-to-next-to-leading-order},''
\href{http://arxiv.org/abs/1805.07243}{{\ttfamily arXiv:1805.07243 [hep-ph]}}.

\bibitem{Echevarria:2014xaa}
M.~G. Echevarria, A.~Idilbi, Z.-B. Kang, and I.~Vitev, ``{QCD Evolution of the
  Sivers Asymmetry},'' \href{http://dx.doi.org/10.1103/PhysRevD.89.074013}{{\em
  Phys. Rev.} {\bfseries D89} (2014) 074013},
\href{http://arxiv.org/abs/1401.5078}{{\ttfamily arXiv:1401.5078 [hep-ph]}}.

\bibitem{DAlesio:2014mrz}
U.~D'Alesio, M.~G. Echevarria, S.~Melis, and I.~Scimemi, ``{Non-perturbative
  QCD effects in $q_{T}$ spectra of Drell-Yan and Z-boson production},''
  \href{http://dx.doi.org/10.1007/JHEP11(2014)098}{{\em JHEP} {\bfseries 11}
  (2014) 098},
\href{http://arxiv.org/abs/1407.3311}{{\ttfamily arXiv:1407.3311 [hep-ph]}}.

\bibitem{Bacchetta:2017gcc}
A.~Bacchetta, F.~Delcarro, C.~Pisano, M.~Radici, and A.~Signori, ``{Extraction
  of partonic transverse momentum distributions from semi-inclusive
  deep-inelastic scattering, Drell-Yan and Z-boson production},''
  \href{http://dx.doi.org/10.1007/JHEP06(2017)081}{{\em JHEP} {\bfseries 06}
  (2017) 081},
\href{http://arxiv.org/abs/1703.10157}{{\ttfamily arXiv:1703.10157 [hep-ph]}}.

\bibitem{Scimemi:2017etj}
I.~Scimemi and A.~Vladimirov, ``{Analysis of vector boson production within TMD
  factorization},''
  \href{http://dx.doi.org/10.1140/epjc/s10052-018-5557-y}{{\em Eur. Phys. J.}
  {\bfseries C78} no.~2, (2018) 89},
\href{http://arxiv.org/abs/1706.01473}{{\ttfamily arXiv:1706.01473 [hep-ph]}}.

\bibitem{Aybat:2011ta}
S.~M. Aybat, A.~Prokudin, and T.~C. Rogers, ``{Calculation of TMD Evolution for
  Transverse Single Spin Asymmetry Measurements},''
  \href{http://dx.doi.org/10.1103/PhysRevLett.108.242003}{{\em Phys. Rev.
  Lett.} {\bfseries 108} (2012) 242003},
\href{http://arxiv.org/abs/1112.4423}{{\ttfamily arXiv:1112.4423 [hep-ph]}}.

\bibitem{Anselmino:2012aa}
M.~Anselmino, M.~Boglione, and S.~Melis, ``{A Strategy towards the extraction
  of the Sivers function with TMD evolution},''
  \href{http://dx.doi.org/10.1103/PhysRevD.86.014028}{{\em Phys. Rev.}
  {\bfseries D86} (2012) 014028},
\href{http://arxiv.org/abs/1204.1239}{{\ttfamily arXiv:1204.1239 [hep-ph]}}.

\bibitem{Kang:2015msa}
Z.-B. Kang, A.~Prokudin, P.~Sun, and F.~Yuan, ``{Extraction of Quark
  Transversity Distribution and Collins Fragmentation Functions with QCD
  Evolution},'' \href{http://dx.doi.org/10.1103/PhysRevD.93.014009}{{\em Phys.
  Rev.} {\bfseries D93} no.~1, (2016) 014009},
\href{http://arxiv.org/abs/1505.05589}{{\ttfamily arXiv:1505.05589 [hep-ph]}}.

\bibitem{Aad:2014xaa}
{\bfseries ATLAS} Collaboration, G.~Aad {\em et~al.}, ``{Measurement of the
  $Z/\gamma^*$ boson transverse momentum distribution in $pp$ collisions at
  $\sqrt{s}$ = 7 TeV with the ATLAS detector},''
  \href{http://dx.doi.org/10.1007/JHEP09(2014)145}{{\em JHEP} {\bfseries 09}
  (2014) 145},
\href{http://arxiv.org/abs/1406.3660}{{\ttfamily arXiv:1406.3660 [hep-ex]}}.

\bibitem{Aad:2015auj}
{\bfseries ATLAS} Collaboration, G.~Aad {\em et~al.}, ``{Measurement of the
  transverse momentum and $\phi ^*_{\eta }$ distributions of Drell-Yan lepton
  pairs in proton-proton collisions at $\sqrt{s}=8$ TeV with the ATLAS
  detector},'' \href{http://dx.doi.org/10.1140/epjc/s10052-016-4070-4}{{\em
  Eur. Phys. J.} {\bfseries C76} no.~5, (2016) 291},
\href{http://arxiv.org/abs/1512.02192}{{\ttfamily arXiv:1512.02192 [hep-ex]}}.

\bibitem{Aaij:2015gna}
{\bfseries LHCb} Collaboration, R.~Aaij {\em et~al.}, ``{Measurement of the
  forward $Z$ boson production cross-section in $pp$ collisions at $\sqrt{s}=7$
  TeV},'' \href{http://dx.doi.org/10.1007/JHEP08(2015)039}{{\em JHEP}
  {\bfseries 08} (2015) 039},
\href{http://arxiv.org/abs/1505.07024}{{\ttfamily arXiv:1505.07024 [hep-ex]}}.

\bibitem{Aaij:2015zlq}
{\bfseries LHCb} Collaboration, R.~Aaij {\em et~al.}, ``{Measurement of forward
  W and Z boson production in $pp$ collisions at $ \sqrt{s}=8 $ TeV},''
  \href{http://dx.doi.org/10.1007/JHEP01(2016)155}{{\em JHEP} {\bfseries 01}
  (2016) 155},
\href{http://arxiv.org/abs/1511.08039}{{\ttfamily arXiv:1511.08039 [hep-ex]}}.

\bibitem{Aaij:2016mgv}
{\bfseries LHCb} Collaboration, R.~Aaij {\em et~al.}, ``{Measurement of the
  forward Z boson production cross-section in pp collisions at $\sqrt{s} = 13$
  TeV},'' \href{http://dx.doi.org/10.1007/JHEP09(2016)136}{{\em JHEP}
  {\bfseries 09} (2016) 136},
\href{http://arxiv.org/abs/1607.06495}{{\ttfamily arXiv:1607.06495 [hep-ex]}}.

\bibitem{Khachatryan:2016nbe}
{\bfseries CMS} Collaboration, V.~Khachatryan {\em et~al.}, ``{Measurement of
  the transverse momentum spectra of weak vector bosons produced in
  proton-proton collisions at $ \sqrt{s}=8 $ TeV},''
  \href{http://dx.doi.org/10.1007/JHEP02(2017)096}{{\em JHEP} {\bfseries 02}
  (2017) 096},
\href{http://arxiv.org/abs/1606.05864}{{\ttfamily arXiv:1606.05864 [hep-ex]}}.

\bibitem{Accardi:2012qut}
A.~Accardi {\em et~al.}, ``{Electron Ion Collider: The Next QCD Frontier},''
  \href{http://dx.doi.org/10.1140/epja/i2016-16268-9}{{\em Eur. Phys. J.}
  {\bfseries A52} no.~9, (2016) 268},
\href{http://arxiv.org/abs/1212.1701}{{\ttfamily arXiv:1212.1701 [nucl-ex]}}.

\bibitem{Dudek:2012vr}
J.~Dudek {\em et~al.}, ``{Physics Opportunities with the 12 GeV Upgrade at
  Jefferson Lab},'' \href{http://dx.doi.org/10.1140/epja/i2012-12187-1}{{\em
  Eur. Phys. J.} {\bfseries A48} (2012) 187},
\href{http://arxiv.org/abs/1208.1244}{{\ttfamily arXiv:1208.1244 [hep-ex]}}.

\bibitem{Brodsky:2012vg}
S.~J. Brodsky, F.~Fleuret, C.~Hadjidakis, and J.~P. Lansberg, ``{Physics
  Opportunities of a Fixed-Target Experiment using the LHC Beams},''
  \href{http://dx.doi.org/10.1016/j.physrep.2012.10.001}{{\em Phys. Rept.}
  {\bfseries 522} (2013) 239--255},
\href{http://arxiv.org/abs/1202.6585}{{\ttfamily arXiv:1202.6585 [hep-ph]}}.

\bibitem{Kikola:2017hnp}
D.~Kikola, M.~G. Echevarria, C.~Hadjidakis, J.-P. Lansberg, C.~Lorcé,
  L.~Massacrier, C.~M. Quintans, A.~Signori, and B.~Trzeciak, ``{Feasibility
  Studies for Single Transverse-Spin Asymmetry Measurements at a Fixed-Target
  Experiment Using the LHC Proton and Lead Beams (AFTER@LHC)},''
  \href{http://dx.doi.org/10.1007/s00601-017-1299-x}{{\em Few Body Syst.}
  {\bfseries 58} no.~4, (2017) 139},
\href{http://arxiv.org/abs/1702.01546}{{\ttfamily arXiv:1702.01546 [hep-ex]}}.

\bibitem{Hadjidakis:2018ifr}
C.~Hadjidakis {\em et~al.}, ``{A Fixed-Target Programme at the LHC: Physics
  Case and Projected Performances for Heavy-Ion, Hadron, Spin and Astroparticle
  Studies},''
\href{http://arxiv.org/abs/1807.00603}{{\ttfamily arXiv:1807.00603 [hep-ex]}}.

\bibitem{Roth:2004ti}
M.~Roth and S.~Weinzierl, ``{QED corrections to the evolution of parton
  distributions},''
  \href{http://dx.doi.org/10.1016/j.physletb.2004.04.009}{{\em Phys. Lett.}
  {\bfseries B590} (2004) 190--198},
\href{http://arxiv.org/abs/hep-ph/0403200}{{\ttfamily arXiv:hep-ph/0403200
  [hep-ph]}}.

\bibitem{Martin:2004dh}
A.~D. Martin, R.~G. Roberts, W.~J. Stirling, and R.~S. Thorne, ``{Parton
  distributions incorporating QED contributions},''
  \href{http://dx.doi.org/10.1140/epjc/s2004-02088-7}{{\em Eur. Phys. J.}
  {\bfseries C39} (2005) 155--161},
\href{http://arxiv.org/abs/hep-ph/0411040}{{\ttfamily arXiv:hep-ph/0411040
  [hep-ph]}}.

\bibitem{Ball:2013hta}
{\bfseries NNPDF} Collaboration, R.~D. Ball, V.~Bertone, S.~Carrazza,
  L.~Del~Debbio, S.~Forte, A.~Guffanti, N.~P. Hartland, and J.~Rojo, ``{Parton
  distributions with QED corrections},''
  \href{http://dx.doi.org/10.1016/j.nuclphysb.2013.10.010}{{\em Nucl. Phys.}
  {\bfseries B877} (2013) 290--320},
\href{http://arxiv.org/abs/1308.0598}{{\ttfamily arXiv:1308.0598 [hep-ph]}}.

\bibitem{deFlorian:2015ujt}
D.~de~Florian, G.~F.~R. Sborlini, and G.~Rodrigo, ``{QED corrections to the
  Altarelli-Parisi splitting functions},''
  \href{http://dx.doi.org/10.1140/epjc/s10052-016-4131-8}{{\em Eur. Phys. J.}
  {\bfseries C76} no.~5, (2016) 282},
\href{http://arxiv.org/abs/1512.00612}{{\ttfamily arXiv:1512.00612 [hep-ph]}}.

\bibitem{deFlorian:2016gvk}
D.~de~Florian, G.~F.~R. Sborlini, and G.~Rodrigo, ``{Two-loop QED corrections
  to the Altarelli-Parisi splitting functions},''
  \href{http://dx.doi.org/10.1007/JHEP10(2016)056}{{\em JHEP} {\bfseries 10}
  (2016) 056},
\href{http://arxiv.org/abs/1606.02887}{{\ttfamily arXiv:1606.02887 [hep-ph]}}.

\bibitem{Mottaghizadeh:2017vef}
M.~Mottaghizadeh, F.~Taghavi~Shahri, and P.~Eslami, ``{Analytical solutions of
  the QED$\otimes$QCD DGLAP evolution equations based on the Mellin transform
  technique},'' \href{http://dx.doi.org/10.1016/j.physletb.2017.08.049}{{\em
  Phys. Lett.} {\bfseries B773} (2017) 375--384},
\href{http://arxiv.org/abs/1707.00108}{{\ttfamily arXiv:1707.00108 [hep-ph]}}.

\bibitem{deFlorian:2018wcj}
D.~de~Florian, M.~Der, and I.~Fabre, ``{QCD$\oplus$QED NNLO corrections to
  Drell Yan production},''
\href{http://arxiv.org/abs/1805.12214}{{\ttfamily arXiv:1805.12214 [hep-ph]}}.

\bibitem{Cieri:2018sfk}
L.~Cieri, G.~Ferrera, and G.~F.~R. Sborlini, ``{Combining QED and QCD
  transverse-momentum resummation for Z boson production at hadron
  colliders},''
\href{http://arxiv.org/abs/1805.11948}{{\ttfamily arXiv:1805.11948 [hep-ph]}}.

\bibitem{GarciaEchevarria:2011md}
M.~Garcia-Echevarria, A.~Idilbi, and I.~Scimemi, ``{SCET, Light-Cone Gauge and
  the T-Wilson Lines},''
  \href{http://dx.doi.org/10.1103/PhysRevD.84.011502}{{\em Phys. Rev.}
  {\bfseries D84} (2011) 011502},
\href{http://arxiv.org/abs/1104.0686}{{\ttfamily arXiv:1104.0686 [hep-ph]}}.

\bibitem{Echevarria:2015uaa}
M.~G. Echevarria, T.~Kasemets, P.~J. Mulders, and C.~Pisano, ``{QCD evolution
  of (un)polarized gluon TMDPDFs and the Higgs $q_T$-distribution},''
  \href{http://dx.doi.org/10.1007/JHEP07(2015)158,
  10.1007/JHEP05(2017)073}{{\em JHEP} {\bfseries 07} (2015) 158},
  \href{http://arxiv.org/abs/1502.05354}{{\ttfamily arXiv:1502.05354
  [hep-ph]}}.
[Erratum: JHEP05,073(2017)].

\bibitem{Moch:2005id}
S.~Moch, J.~A.~M. Vermaseren, and A.~Vogt, ``{The Quark form-factor at higher
  orders},'' \href{http://dx.doi.org/10.1088/1126-6708/2005/08/049}{{\em JHEP}
  {\bfseries 08} (2005) 049},
\href{http://arxiv.org/abs/hep-ph/0507039}{{\ttfamily arXiv:hep-ph/0507039
  [hep-ph]}}.

\bibitem{Moch:2004pa}
S.~Moch, J.~A.~M. Vermaseren, and A.~Vogt, ``{The Three loop splitting
  functions in QCD: The Nonsinglet case},''
  \href{http://dx.doi.org/10.1016/j.nuclphysb.2004.03.030}{{\em Nucl. Phys.}
  {\bfseries B688} (2004) 101--134},
\href{http://arxiv.org/abs/hep-ph/0403192}{{\ttfamily arXiv:hep-ph/0403192
  [hep-ph]}}.

\bibitem{Moch:2017uml}
S.~Moch, B.~Ruijl, T.~Ueda, J.~A.~M. Vermaseren, and A.~Vogt, ``{Four-Loop
  Non-Singlet Splitting Functions in the Planar Limit and Beyond},''
  \href{http://dx.doi.org/10.1007/JHEP10(2017)041}{{\em JHEP} {\bfseries 10}
  (2017) 041},
\href{http://arxiv.org/abs/1707.08315}{{\ttfamily arXiv:1707.08315 [hep-ph]}}.

\bibitem{Collins:2014jpa}
J.~Collins and T.~Rogers, ``{Understanding the large-distance behavior of
  transverse-momentum-dependent parton densities and the Collins-Soper
  evolution kernel},'' \href{http://dx.doi.org/10.1103/PhysRevD.91.074020}{{\em
  Phys. Rev.} {\bfseries D91} no.~7, (2015) 074020},
\href{http://arxiv.org/abs/1412.3820}{{\ttfamily arXiv:1412.3820 [hep-ph]}}.

\bibitem{Scimemi:2018xaf}
I.~Scimemi and A.~Vladimirov, ``{Systematic analysis of double-scale
  evolution},''
\href{http://arxiv.org/abs/1803.11089}{{\ttfamily arXiv:1803.11089 [hep-ph]}}.

\bibitem{Echevarria:2015usa}
M.~G. Echevarria, I.~Scimemi, and A.~Vladimirov, ``{Transverse momentum
  dependent fragmentation function at next-to--next-to--leading order},''
  \href{http://dx.doi.org/10.1103/PhysRevD.93.011502,
  10.1103/PhysRevD.94.099904}{{\em Phys. Rev.} {\bfseries D93} no.~1, (2016)
  011502}, \href{http://arxiv.org/abs/1509.06392}{{\ttfamily arXiv:1509.06392
  [hep-ph]}}.
[Erratum: Phys. Rev.D94,no.9,099904(2016)].

\bibitem{Kilgore:2011pa}
W.~B. Kilgore and C.~Sturm, ``{Two-Loop Virtual Corrections to Drell-Yan
  Production at order $\alpha_s \alpha^3$},''
  \href{http://dx.doi.org/10.1103/PhysRevD.85.033005}{{\em Phys. Rev.}
  {\bfseries D85} (2012) 033005},
\href{http://arxiv.org/abs/1107.4798}{{\ttfamily arXiv:1107.4798 [hep-ph]}}.

\bibitem{Kilgore:2013uta}
W.~B. Kilgore, ``{The Two-Loop Infrared Structure of Amplitudes with Mixed
  Gauge Groups},'' \href{http://dx.doi.org/10.1140/epjc/s10052-013-2603-7}{{\em
  Eur. Phys. J.} {\bfseries C73} (2013) 2603},
\href{http://arxiv.org/abs/1308.1055}{{\ttfamily arXiv:1308.1055 [hep-ph]}}.

\bibitem{Echevarria:2016mrc}
M.~G. Echevarria, A.~Idilbi, K.~Kanazawa, C.~Lorcé, A.~Metz, B.~Pasquini, and
  M.~Schlegel, ``{Proper definition and evolution of generalized transverse
  momentum dependent distributions},''
  \href{http://dx.doi.org/10.1016/j.physletb.2016.05.086}{{\em Phys. Lett.}
  {\bfseries B759} (2016) 336--341},
\href{http://arxiv.org/abs/1602.06953}{{\ttfamily arXiv:1602.06953 [hep-ph]}}.

\end{thebibliography}\endgroup

\end{document}